\documentclass[10pt]{iopart}
\usepackage{iopams}

\newcommand{\bra}{\left\langle}
\newcommand{\ket}{\right\rangle}

\newcommand{\pder}[2]{\frac{\partial #1}{\partial  #2}}
\newcommand{\pdert}[2]{\frac{\partial^2 #1}{\partial  #2^2}}
\newcommand{\der}[2]{\frac{d #1}{d  #2}}

\newcommand{\vs}{v_{\rm s}}
\newcommand{\ps}{p_{\rm st}}

\newcommand{\ve}{\varepsilon}
\newcommand{\fp}{f_{\rm p}}
\newcommand{\Kp}{K_{\perp}}

\newcommand{\LFP}{\mathcal{L}_\mathrm{FP}}
\newcommand{\tLFP}{\mathcal{L}_\mathrm{FP}^\ddagger}

\begin{document}
% Journal identifier can be put here if required, e.g.
\jl{1}

\title[Fluctuating response equation from a Langevin equation]{Exact transformation of a Langevin equation to a fluctuating response equation}

\author{Takahiro Harada\dag\footnote[3]
{To whom correspondence should be addressed
(harada@chem.scphys.kyoto-u.ac.jp).}, Kumiko Hayashi\ddag\ and 
Shin-ichi Sasa\ddag}

\address{\dag\ Department of Physics, Graduate School of Science, Kyoto University, 
Kyoto 606-8502, Japan}

\address{\ddag\ Department of Pure and Applied Sciences, University of Tokyo, 
Komaba, Tokyo 153-8902, Japan}

\begin{abstract}
We demonstrate that a Langevin equation that describes the motion of a Brownian particle under non-equilibrium conditions can be exactly transformed to a special equation that explicitly exhibits the response of the velocity to a time dependent perturbation.
This transformation is constructed on the basis of an operator formulation 
originally used  in nonlinear perturbation theory for differential equations by extending it to stochastic analysis.
We find that the obtained expression is useful for the calculation of fundamental quantities of the system, and that it provides a physical basis for the decomposition of the forces in the Langevin description into effective driving, dissipative, and random forces in a large-scale description. 
\end{abstract}


\submitted

%\maketitle

%%%%%%%%%%%%%% main text starts %%%%%%%%%%%%%%%%%%%

\section{Introduction}
\label{intro}

Owing to technological advances in the methods of manipulating systems on  sub-micrometer length scales in aqueous solution, there is 
increased interest in studying the nonequilibrium nature of such small systems. 
Recent studies have yielded several universal relations, including the fluctuation theorem \cite{ft}, the Jarzynski equality \cite{jar} and the Hatano-Sasa nonequilibrium steady state equality \cite{hata}, and the validity of these relations has been verified experimentally in small systems consisting of beads and bead-RNA complexes \cite{eft,ejar,ehata}.
In addition to the verification of these equalities, there is an interesting experimental study providing new information concerning the nature of nonequilibrium systems, in which the velocity response to an external perturbation and the correlation of the corresponding fluctuations were measured in a bead system \cite{harada1}. We should also mention that there has been a substantial progress in understanding stochastic resonance \cite{sr, srex}.

Although great progress has been made in the experimental investigation of such small systems, direct measurements have been possible only for kinematic quantities, such as the position of a bead. 
Given this situation, in order to elucidate the mechanics of small systems, it is necessary to extract information regarding mechanical quantities, such as `force' 
and `energy', from kinematic quantities.
However, this task is not easily accomplished in general. For instance, consider the problem of guessing the equation of motion of a motor protein on the basis of experimental results.
It can then be understood why it is difficult to determine an effective potential for the center of mass and to evaluate the size of dissipation effects, because neither a canonical distribution nor a fluctuation-response relation exist for nonequilibrium systems.
(By contrast, these can be utilized to determine an effective potential and a dissipation strength near equilibrium states.)

In this paper, we study a nonequilibrium Langevin system for the purpose of developing a new theoretical method to extract information regarding mechanical quantities in small systems.
Specifically, we attempt to re-express a Langevin equation in terms of directly measurable quantities.
Here, let us recall that for differential equations, there exists a theoretical method by which, for example, a Rayleigh equation that describes nonlinear oscillations can be transformed perturbatively into  a simple equation displaying the observed frequency \cite{pbook}.
Following this idea, it might be useful if we could transform a Langevin equation into a special form that makes mechanical properties manifest.
Specifically, motivated by a recent phenomenological study of energy efficiency \cite{Harada}, we regard the response to a time-dependent perturbative force as a key property that connects kinematic and mechanical quantities.

Here, we point out that a Langevin equation is equivalent to the corresponding Fokker-Planck equation as long as we  consider statistical properties. 
However, when we study problems related to force and energy, analysis of a Langevin equation is more appropriate, because it can be regarded as an equation of force balance. 
Furthermore, physical quantities for  a single trajectory, which are often observed in bio-mechanical systems \cite{kine}, can be described only by a Langevin equation.
For these reasons, we seek a useful representation of a Langevin equation rather than a  Fokker-Planck equation. 

%%% model %%%

Explicitly, we study the simple one-dimensional Langevin equation
\begin{equation}
\gamma \dot x(t)= f-U'(x(t)) + \xi(t)+ \ve \fp(t).
\label{eq:Langevin}
\end{equation}
Here, $\gamma$ is a friction constant, $f$ is a constant driving force, $U(x)$ is a periodic potential of period $\ell$, and the prime denotes differentiation. 
Also, $\xi(t)$ is  Gaussian white noise satisfying 
\begin{equation}
\bra \xi(t) \xi(s) \ket = 2\gamma T \delta(t-s),
\label{noise}
\end{equation}
where $\bra \cdots \ket$ denotes the average over all noise histories.  
(See the references \cite{Risken} and \cite{Reimann} for discussion of the 
physical basis of this equation in the case $\ve=$0.)  
The term $\ve \fp(t) $ with sufficiently small $\ve$ represents a perturbation force we use to investigate the response of the system. 
An initial condition $x(t_0)=x_0$ is given at $t=t_0$, and we often take the limit $t_0 \to -\infty$.
In the argument below, 
we denote the average over all noise histories in  the limit $t_0 \to -\infty$ as $\bra \cdots \ket_\ve $.
Under these conditions, there exists the time-dependent distribution function $p_\infty (\theta, t)$, such that the relation,
\begin{equation}
\bra A(x(t))\ket_\ve=\int_0^\ell d \theta A(\theta) p_\infty(\theta,t),
\label{pin}
\end{equation}
is satisfied for any $\ell$-periodic function $A(\theta)$.
Note that $p_\infty (\theta,t)$ can be expanded in the form
\begin{equation}
p_\infty(\theta,t)=\ps(\theta)+\ve p^{(1)}(\theta,t)+O(\ve^2).
\label{pinex}
\end{equation}

%%% theorem %%%
With the above preparation, in Section \ref{sec:proof1}, we  prove the following theorem.
\vskip3mm
\noindent
{\bf Theorem :} Under the condition $t_0 \to -\infty$, the equation  (\ref{eq:Langevin}) can be transformed into the form
\begin{equation}
\int_0^{\infty} ds L(s)(\dot x(t-s)-\vs)
=\int_0^{\infty} ds L(s)\zeta(t-s)
+\xi(t)+\ve \fp(t),
\label{norm:1}
\end{equation}
where $\vs\equiv \bra \dot x (t) \ket_0$ is the steady state velocity.
The function $L(s)$ here is determined by the system parameters $\gamma$, $f$, $U(x)$ and $T$, and is independent of $\ve$.
$L(s)$ is determined so that it satisfies the causality property, i.e., $L(t) = 0$ for $t < 0$.
The function $\zeta(t)$ depends on $\{ x(s) \}_{t_0 \le s \le t}$, $\{\xi(s) \}_{t_0 \le s \le t}$, and $\{\fp(s) \}_{t_0 \le s \le t} $, and most importantly it satisfies the relation
\begin{equation}
\bra \zeta(t)\ket_\ve = O(\ve^2).
\label{norm:2}
\end{equation}
These functions, $L(s)$ and $\zeta(s)$, are determined from the set of eigenfunctions of the Fokker-Planck operator corresponding to the Langevin equation (\ref{eq:Langevin}), as explained in Section \ref{sec:key}. 

We now explain the physical significance of this theorem. 
First, note that (\ref{norm:1}) is equivalent to the form
\begin{equation}
\dot x(t)-\vs=\zeta(t)+\int_0^{\infty} ds R(s) (\xi(t-s)+\ve \fp(t-s)).
\label{norm:3}
\end{equation}
Here, the quantity $R(t)$ is determined from $L(t)$ through the relation
\begin{equation}
\tilde R(\omega)=\tilde L(\omega)^{-1},
\label{LRcorrespondence}
\end{equation}
where $\tilde R (\omega)$ and $\tilde L (\omega)$ are the Fourier transforms of $R(t)$ and $L(t)$.
$R(t)$ for $t>0$ is obtained by the inverse Fourier transform and we also require the causality as $R(t) =0$ for $t < 0$.
Throughout this paper, we use similar notation, with the Fourier transform of an arbitrary function $A(t)$ given by
\begin{equation}
\tilde A(\omega)=\int_{-\infty}^\infty dt \e^{i\omega t } A(t).
\end{equation}
Then, the average of (\ref{norm:3}) becomes
\begin{equation}
\bra \dot x(t) \ket_\ve -\vs= \ve \int_0^{\infty}ds R(s)\fp(t-s)+O(\ve^2).
\label{resdef0}
\end{equation}
This equation characterizes the linear response of the velocity to the perturbation.
We therefore call $R(s)$ a ``response function''.
Its Fourier transform, $\tilde R(\omega)$, is called the ``dynamic susceptibility''.
The expression (\ref{norm:3}), which is equivalent to the original equation of motion (\ref{eq:Langevin}),  explicitly represents the response to a perturbation.

%%% effective %%%

Next, we show how the expression (\ref{norm:1}) is related to the problem of force decomposition in a large-scale description. 
Let us express the force acting on the particle at time $t$ as
\begin{equation}
\phi(t) \equiv f-U'(x(t))
\label{pdef}
\end{equation}
and consider a finite time average of $\phi(t)$.
For convenience, we express the finite time average of a quantity $A(t)$ as
\begin{equation}
\bar A_\tau(t)=\frac{1}{\tau} \int_t^{t+\tau} ds A(s).
\label{finav}
\end{equation}
From (\ref{eq:Langevin}), (\ref{norm:1}), (\ref{pdef}) and (\ref{finav}), we find
\begin{eqnarray}
\bar \phi_\tau (t) &=& \gamma \frac{x(t+\tau)-x(t)}{\tau}+ \int_0^{\infty} ds L(s) \bar \zeta_\tau(t-s)\nonumber \\
& & - \int_0^{\infty} ds L(s)\left(
\frac{x(t-s+\tau)-x(t-s)}{\tau}-\vs \right).
\label{p6-a}
\end{eqnarray}
We now assume that $\tau$ is sufficiently larger than the inverse of the smallest decay rate of $L(s)$.
Then, because $\bar A_\tau (t)$ varies slowly with respect to $t$, (\ref{p6-a}) becomes
\begin{equation}
\bar \phi_\tau (t) \simeq   \tilde L(0) \vs -(\tilde L(0)-\gamma) 
\frac{x(t+\tau)-x(t)}{\tau} +  \tilde L(0)\bar \zeta_\tau(t).
\label{p6-b}
\end{equation}
This expression implies that a time averaged force $\bar \phi_\tau (t)$ can be decomposed into a driving force (the first term), a dissipative force (the second term), and a random force (the third term).
In the reference \cite{HSV}, the force decomposition for a large-scale description with (\ref{eq:Langevin}) is investigated, and it is shown that the condition 
\begin{equation}
\lim_{\tau \to \infty} \tau \bra \bar \zeta_\tau \bar \xi_\tau  \ket_0 =0
\label{HScon}
\end{equation}
uniquely determines the force decomposition of the type given in (\ref{p6-b}). 
In other words, only the component orthogonal to $\bar \xi_\tau$ is regarded as a random component of $\bar \phi_\tau$ in such a description. 
In Section \ref{sec:remark}, we confirm that this orthogonality condition is satisfied. 
Actually, we have arrived at the expression (\ref{norm:1}) by seeking a definition of $\zeta(t)$ for which (\ref{HScon}) holds.
In this sense, (\ref{HScon}) was the guiding principle used in obtaining the transformation yielding (\ref{norm:1}) from (\ref{eq:Langevin}). 

In Section \ref{sec:proof2}, we demonstrate that the expression (\ref{norm:3}) is useful for calculating fundamental statistical quantities.
As already stated, the susceptibility $\tilde R(\omega)$ is obtained as the Fourier transform of $R(t)$.
In \ref{app:1}, we provide a consistency check of our theory by comparing this quantity $\tilde R(\omega)$ with an expression that is obtained directly from the definition of the dynamic susceptibility.
Furthermore, the Fourier transform $\tilde C(\omega)$ of the time correlation of velocity fluctuations,
\begin{equation}
C(t) \equiv  \bra (\dot x(t)-\vs) (\dot x(0)-\vs) \ket_0,
\label{ctdef}
\end{equation}
can also be calculated directly from (\ref{norm:3}).
In particular, for the case $\omega=0$, we can derive the following formulae:
\begin{eqnarray}
\tilde R(0) 
&=& 
\frac{\ell}{\gamma} \frac{\int_0^\ell d\theta I_-(\theta)I_+(\theta)}
{\left( \int_0^\ell dx I_-(\theta) \right)^2},
\label{mud}
\\ 
\frac{\tilde C(0)}{2}
&=&   
\frac{T}{\gamma} \ell^2
\frac{\int_0^\ell d\theta I_-(\theta) I_+(\theta)^2}
{\left(\int_0^\ell d\theta I_-(\theta) \right)^3},
\label{diff}
\end{eqnarray}
where
\begin{equation}
I_{\pm}(\theta)=\int_0^\ell d\theta' e^{\pm \beta U(\theta)
\mp \beta U(\theta \mp \theta')-\beta f \theta'},
\end{equation}
with $\beta \equiv 1/T$.
The key technical lemma used in the derivations of the above relations is proved in \ref{app:2}.

We  note that $\tilde R(0)$ is equal to the differential mobility, $d\vs/df$, and that the expression (\ref{mud}) was calculated from the functional form of $\vs(f)$ given in the reference \cite{HSIII}. 
Also, $\tilde C(0)/2$ is equal to the diffusion constant $D$ defined by
\begin{equation}
D\equiv \lim_{t \to \infty} \frac{1}{2t}
\bra (x(t)-x(0)-\vs t)^2 \ket_0 .
\label{Ddef}
\end{equation}
The expression  (\ref{diff}) was first derived in the reference \cite{Reimann} employing a recursion formula that is established in the theory of stochastic processes.
Subsequently, the same expression was obtained in a perturbation theory treatment applied to the Fokker-Planck equation corresponding to (\ref{eq:Langevin}) \cite{HSIII}. 
The method of derivation used in the present work is different from those used in the previous studies.

Furthermore, from the correspondence of $\tilde R(0)$ with $d\vs/df$ and from the definition of $D$ in (\ref{Ddef}), the nature of each term in the force decomposition (\ref{p6-b}) is clearly understood.
By substituting (\ref{p6-b}) into the $\tau$-averaged form of 
(\ref{eq:Langevin}), we obtain a large-scale description of the Langevin equation:
\begin{equation}
\tilde L (0) \frac{x(t+\tau) - x(t)}{\tau} = \tilde L(0) \vs + \bar \xi_\tau (t) + \tilde L(0) \bar \zeta_\tau (t).
\label{effLangevin}
\end{equation}
The effective dissipation coefficient, $\tilde L(0)$, is the inverse of the differential mobility $\tilde R(0)$ as explained in (\ref{LRcorrespondence}).
The magnitude of the fluctuating forces, $\langle (\bar \xi_\tau (t) + \tilde L(0) \bar \zeta_\tau (t) )^2 \rangle_0$ is expressed in terms of the diffusion constant as $2 D /\tilde R(0)^2$, which provides the renormalization of the bare noise intensity $2 \gamma T$.
These quantities are obtained by use of the explicit expressions given in (\ref{mud}) and (\ref{diff}).

%%%%%%%%%%%%%%%%%%%%%%%%%%%%%%%%%%%%%%%%%%%%%%%%%%%%%%%%%%%%
%%%%%%%%%%%%%%%%%%%%%%%%%%%%%%%%%%%%%%%%%%%%%%%%%%%%%%%%%%%%

\section{Proof of Theorem}
\label{sec:proof1}

\subsection{Operator formulation}

In order to make the argument mathematically clearer, we express the Langevin equation (\ref{eq:Langevin})  as 
\begin{equation}
\gamma dx(t) = \left\{ f- U'(x(t))  \right\}dt 
+ \sqrt{2\gamma T} dw(t) +\ve \fp(t)dt,
\label{ap:ItoLangevin}
\end{equation}
where $w(t)$ represents a Wiener process \cite{sabook}.
Let $A(x)$ be an arbitrary function of $x$. Then, It\^o's formula gives
\begin{eqnarray}
dA(x(t)) &=&  \frac{F(x(t))}{\gamma} A'(x(t))dt
        +\frac{T}{\gamma}A''(x(t)) dt \nonumber \\
& & +\sqrt{\frac{2T}{\gamma}}A'(x(t)) \cdot dw
+\frac{1}{\gamma} \ve \fp(t) A'(x(t)) dt,
\label{ito-app}
\end{eqnarray}
where
\begin{equation}
F(x) \equiv f-U'(x).
\label{Fdef}
\end{equation}
Here, $\cdot$ represents the product in the It\^o interpretation.
In conventional notation, this is written as 
\begin{equation}
\der{A(x(t))}{t}= \left. \Lambda A(x) \right|_{x=x(t)} + A'(x(t)) \cdot \frac{1}{\gamma}(\xi(t) + \ve \fp(t)),
\label{eq:dynamics}
\end{equation}
where 
\begin{equation}
\Lambda \equiv \frac{F(x)}{\gamma} \pder{}{x}
+\frac{T}{\gamma}\pdert{}{x}.
\label{eq:Lambda}
\end{equation}
We can express the solution of (\ref{eq:dynamics}) in the form
\begin{equation}
A(x(t)) = \left. \mathcal{G}(t) A(x) \right|_{x=x_0},
\label{def-g}
\end{equation}
by introducing a time-dependent operator, $\mathcal{G}(t)$, 
that does not depend on $A(x)$.
Substituting (\ref{def-g}) into (\ref{eq:dynamics}), 
we obtain an equation for $\mathcal{G}(t)$ as
\begin{equation}
\der{\mathcal{G}(t)}{t} = \mathcal{G} (t) \Lambda 
+ \mathcal{G}(t) \pder{}{x} \cdot \frac{1}{\gamma} (\xi(t) + \ve \fp(t)).
\label{dyn-g}
\end{equation}
With the initial condition $\mathcal{G}(t_0)=1$,
the formal solution of this equation is derived as
\begin{equation}
\mathcal{G}(t) = 
e^{(t-t_0)\Lambda} + 
\int_0^{t-t_0} ds \mathcal{G}(t-s) \pder{}{x} 
e^{s\Lambda} \cdot \frac{1}{\gamma}(\xi(t-s) + \ve \fp(t-s)).
\label{sol-g}
\end{equation}
Thus, because the force acting on the particle at time $t$ 
is given by
\begin{eqnarray}
\phi(t) &=& F(x(t)) \nonumber \\
        &=& \left. \mathcal{G}(t) F(x) \right|_{x=x_0},
\end{eqnarray}
we obtain 
\begin{eqnarray} 
\phi(t) &=& \left. e^{(t-t_0)\Lambda} F(x) \right|_{x = x_0} \nonumber \\
& & + \int_0^{t-t_0} ds \frac{\xi(t-s) + \ve \fp(t-s)}{\gamma} \cdot \left. \pder{}{x} 
e^{s \Lambda} F(x) \right|_{x=x(t-s)},
\label{eq:g}
\end{eqnarray}
where we have used (\ref{def-g}) in the derivation of  the 
second term on the right-hand side.

In this way, we have reformulated the original nonlinear stochastic differential equation in terms of the operator $\Lambda$.
We note that a theoretical framework for carrying out such a reformulation in the case of differential equations was developed through application of nonlinear perturbation theory \cite{Bogaevski}.
Also, analysis based on the ``microscopic distribution function'' \cite{Mori} is essentially the same as the present formulation. 

\subsection{Preparation of functional space}

In order to make our investigation of (\ref{eq:g}) concrete, we introduce a functional space, $\mathcal{H}$, consisting of all complex valued, square 
integrable, periodic functions of $\theta$ on the interval $0 \le \theta \le \ell$.
We endow this space with the inner product
\begin{equation}
(h_1,h_2)
=\int_0^\ell d\theta h_1^*(\theta)h_2(\theta),
\end{equation}
for $h_1, h_2 \in \mathcal{H}$, where $^*$ denotes complex conjugation.
All the eigenvalues, $-\lambda_j$, and the corresponding eigenfunctions, $\Phi_j (\theta)$, of the operator $\Lambda$ in $\mathcal{H}$ are determined by the equation
\begin{equation}
\Lambda \Phi_j(\theta) = -\lambda_j \Phi_j(\theta),
\label{Phidef}
\end{equation}
where the index $j = 0, \pm 1, \pm 2, \cdots$ is determined by the relations $\lambda_j = \lambda_{-j}^*$ and $\lambda_0 = 0 < \mathrm{Re}(\lambda_{\pm 1}) < \mathrm{Re}(\lambda_{\pm 2}) < \cdots$ holding among their eigenvalues.
When a complex eigenvalue happens to be degenerate, the corresponding labeling is modified appropriately. 

Because $\Lambda$ given in (\ref{eq:Lambda}) is not a Hermitian operator on $\mathcal{H}$, it is convenient to introduce the adjoint operator of $\Lambda$ through the relation
\begin{equation}
(\LFP h_1, h_2) \equiv (h_1,\Lambda h_2),
\end{equation}
where $\LFP$ is the Fokker-Planck operator:
\begin{equation}
\LFP =  - \frac{1}{\gamma}\pder{}{\theta}F(\theta)
+\frac{T}{\gamma} \pdert{}{\theta}.
\end{equation}
Note that the set of eigenvalues of $\LFP$ is identical to that of $\Lambda$. Then, we denote the eigenfunctions of $\LFP$ by $\Psi_j (\theta)$ and choose their labeling so that we have
\begin{equation}
\LFP \Psi_j(\theta) =-\lambda_j^* \Psi_j(\theta).
\end{equation}
We can choose these eigenfunctions such that the following hold:
\begin{eqnarray}
\int_0^\ell d\theta \Psi_i^*(\theta) \Phi_j(\theta) &=&  \delta_{ij}, 
\label{ortho1} \\
\sum_{j=-\infty}^\infty 
\Psi_j^*(\theta) \Phi_j(\theta') &=&  \delta(\theta-\theta').
\end{eqnarray}
In particular, we determine the normalization factor for the zero eigenfunctions 
\begin{eqnarray}
\Psi_0(\theta) &=& \ps(\theta) , \label{zero1} \\
\Phi_0(\theta) &=& 1.
\end{eqnarray}

Because $F(\theta) \in \mathcal{H}$, we can expand $F(\theta)$ in the form
\begin{equation}
F(\theta) = \sum_{j=-\infty}^{\infty}  (\Psi_j,F) \Phi_j(\theta).
\label{exF}
\end{equation}
Here, the following relation is easily confirmed:
\begin{equation}
(\Psi_0,F)=\gamma \vs.
\end{equation}
Then, from (\ref{eq:g}) and (\ref{exF}), we obtain
\begin{eqnarray}
\phi(t) &=& \gamma \vs+b(t) \nonumber \\
& & +\int_0^{t-t_0} ds (\xi(t-s)+\ve \fp(t-s))\cdot K(s,x(t-s)),
\label{p2}
\end{eqnarray}
with 
\begin{equation}
b(t) \equiv \sum_{j=-\infty, j \neq 0}^{\infty} (\Psi_j,F) \e^{-\lambda_j (t-t_0)} \Phi_j (x_0) 
\end{equation}
and 
\begin{equation}
K(s, \theta)\equiv 
\frac{1}{\gamma} \sum_{j=-\infty}^\infty \e^{-s \lambda_j}(\Psi_j,F) \Phi_j'(\theta),
\label{Kdef}
\end{equation}
for $s > 0$ and $K(s, \theta) \equiv 0$ for $s < 0$.
Note that $b(t)$ satisfies the relation
\begin{equation}
\lim_{t_0 \to -\infty} b(t)=0.
\label{bzero}
\end{equation}

\subsection{Key step} \label{sec:key}

We decompose $K(s,\theta)$ into two parts as
\begin{equation}
K(s,\theta)=K_0(s)+\Kp(s,\theta),
\label{Kpdef}
\end{equation}
where we have
\begin{equation}
K_0(s)\equiv (\Psi_0, K(s, \cdot)),
\label{K0def}
\end{equation}
for $s > 0$ and $K_0 (s) \equiv 0$ for $s < 0$.
We choose this decomposition in order to satisfy the relation
\begin{eqnarray}
\bra \Kp(s,x(t)) \ket_0 &=& 
\int_0^\ell d\theta \ps(\theta) \Kp(s,\theta) \nonumber \\
&=& 0.
\label{zerocon}
\end{eqnarray}
Indeed, this equality can be derived from  (\ref{ortho1}) and (\ref{zero1}).  
Now, defining the quantity
\begin{equation}
\zeta(t)\equiv  \frac{1}{\gamma}
\int_0^{t-t_0} ds (\xi(t-s)+\ve \fp(t-s)) \cdot \Kp(s,x(t-s)),
\label{zeta:def}
\end{equation}
we rewrite $\phi(t)$ as 
\begin{equation}
\phi(t) = \gamma \vs+b(t)
+\int_0^{t-t_0} ds K_0(s)(\gamma \dot x(t-s)-\phi(t-s))
+\gamma \zeta(t).
\label{p4}
\end{equation}
Here, we introduce
\begin{eqnarray}
\varphi(t) &\equiv &  \phi(t)-\gamma \vs, \\
\Delta(t)  &\equiv & \dot x(t)-\vs
\end{eqnarray}
and take the limit $t_0 \to -\infty$.
The Fourier transform of (\ref{p4}) then yields 
\begin{equation}
\tilde \varphi(\omega) = \tilde K_0(\omega)
(\gamma \tilde{\Delta}(\omega)-\tilde \varphi(\omega))
+ \gamma \tilde\zeta (\omega).
\label{p5}
\end{equation}
From this, we derive
\begin{equation}
\tilde \varphi(\omega) = \frac{1}{1+\tilde K_0(\omega)}
[\tilde K_0(\omega)\gamma \tilde{\Delta}(\omega)+
\gamma \tilde \zeta (\omega)].
\label{p52}
\end{equation}
The inverse Fourier transform of this expression yields
\begin{equation}
\phi(t) = \gamma \dot x(t)- \int_0^{\infty} ds L(s)(\dot x(t-s)-\vs)
+ \int_0^{\infty} ds L(s)\zeta(t-s),
\label{p6}
\end{equation}
with 
\begin{equation}
L(t)=\int_{-\infty}^\infty \frac{d\omega}{2\pi} \e^{-i\omega t} \frac{\gamma}{1+ \tilde  K_0(\omega)}
\label{Ldef}
\end{equation}
for $t > 0$ and $L(t)=0$ for $t <0$.
Then, recalling (\ref{pdef}), we find that (\ref{p6}) leads to the expression (\ref{norm:1}).

Next, using (\ref{pin}) and (\ref{pinex}), we find that $\zeta(t)$ defined by (\ref{zeta:def}) satisfies
\begin{equation}
\bra \zeta(t)\ket_\ve = \frac{\ve}{\gamma}
\int_0^{\infty} ds \fp(t-s) \bra \Kp(s,x(t-s)) \ket_0 +O(\ve^2),
\label{xir3}
\end{equation}
where we have used the It\^o interpretation. 
Finally, from (\ref{zerocon}), we derive (\ref{norm:2}).

\subsection{Remark}
\label{sec:remark}

As seen in the above discussion, the proper definition of $\zeta$ is essential to obtain  (\ref{norm:2}).
We arrived at the definition (\ref{zeta:def}) in the following way.
Intuitively, $\zeta$ corresponds to what is in some sense the ``random'' part of the force $\phi(t)$ for the case $\ve=0$. 
With this in mind, we assume the following conditions:
\begin{eqnarray}
\bra \zeta(t) \ket_0 = 0,   \label{noise1} \\
\bra \zeta(t) \xi(t') \ket_0 = 0.  \label{noise2}
\end{eqnarray}
The condition (\ref{noise1}) is obviously necessary, and the condition (\ref{noise2}) is inspired by the orthogonality condition (\ref{HScon}) reported in the reference \cite{HSV}.
From the definition (\ref{zeta:def}), it is readily confirmed
\begin{eqnarray}
\bra \zeta(t) \xi(t') \ket_0
&=&
\frac{1}{\gamma}
\int_0^{\infty} ds  \bra \Kp(s,x(t-s)) \ket_0 \bra \xi(t-s)\xi(t') \ket
\label{no2} \nonumber \\
&=& 0,
\label{no4}
\end{eqnarray}
where (\ref{zerocon}) has been used in the last line. It should be noted that the condition (\ref{noise2}) is satisfied regardless of the interpretation of the multiplication between $\zeta(t)$ and $\xi(t)$. Then, guided by the condition (\ref{noise2}), we find the definition  (\ref{zeta:def}), and this leads to the proof of (\ref{norm:2}). 

%%%%%%%%%%%%%%%%%%%%%%%%%%%%%%%%%%%%%%%%%%%%%%%%%%%%%%
%%%%%%%%%%%%%%%%%%%%%%%%%%%%%%%%%%%%%%%%%%%%%%%%%%%%%%

\section{Application}
\label{sec:proof2}

First, we consider the response function $R(t)$. 
From (\ref{Kdef}), (\ref{K0def}) and (\ref{Ldef}), we obtain the following expression of the dynamic susceptibility:
\begin{equation}
\tilde R (\omega) = \frac{1}{\gamma} \left[ 1+\frac{1}{\gamma} 
\sum_{j=-\infty}^\infty \frac{(\Psi_j, F)}{\lambda_j - i \omega} \left( \Psi_0, \Phi'_j \right) \right]
\label{R}
\end{equation}
Replacement of the infinite sum in the right-hand side of (\ref{R}) with a finite sum enables us to calculate $\tilde R(\omega)$ within a required precision because the contribution of the term with large  $|j|$ to the right-hand side in (\ref{R}) becomes on the order of $|j|^{-2}$ for sufficiently large $|j|$.
The response function $R(t)$ is obtained from the inverse Fourier transform.

Next, note that the most important quantity characterizing the statistical properties of fluctuations described by (\ref{eq:Langevin}) might be the velocity correlation function (\ref{ctdef}).
For the equilibrium case, $f=0$, as a result of the detailed balance condition, $C(t)$ is determined by $R(t)$ through the fluctuation-response relation.
However, for nonequilibrium cases, there is no such relation.
Thus, we need to derive an expression of $C(t)$ independently of $R(t)$. 

Substituting (\ref{zeta:def}) into  (\ref{norm:3}), we find that it is convenient to define 
\begin{equation}
M(s,\theta)\equiv \frac{1}{\gamma}\Kp(s,\theta)+R(s).
\label{M:def}
\end{equation}
We then obtain 
\begin{eqnarray}
C(t) = 2 \gamma T  \int_0^\infty ds
\bra M(s,\theta) M(s-t,\theta) \ket_0,
\end{eqnarray}
by use of (\ref{noise}).
Noting that $C(t)=C(-t)$ and that $M(t, \theta)=0$ for $t <0$, we calculate
\begin{eqnarray}
\tilde C(\omega) &=& 
2 \gamma T  
\bra |\tilde M(\omega,\theta) |^2 \ket_0 \nonumber  \\
&=& 
\frac{2 T}{\gamma}  
\left( \Psi_0,  
\left\vert 1+\frac{1}{\gamma} 
\sum_{j=-\infty}^\infty \frac{(\Psi_j, F)}{\lambda_j - i \omega} \Phi'_j \right\vert^2 
\right).
\label{C}
\end{eqnarray}
With this expression, $\tilde C(\omega)$ can be calculated with a required precision for the same reason as in the case of $\tilde R(\omega)$.  
The correlation function $C(t)$ is obtained from the inverse Fourier transform.
We remark that the result (\ref{C}) does not depend on the manner in which we interpret the multiplication between $\dot x(t)$ and $\dot x(0)$ in (\ref{ctdef}).

Here, we derive more compact expressions for $\tilde R(0)$ and $\tilde C(0)$.
First, for later convenience, we define 
\begin{equation}
V(\theta) \equiv  U(\theta)- f \theta.
\end{equation}
Obviously, $F= - V'$. 
Then, it is easy to confirm the equality
\begin{eqnarray}
\lefteqn{\e^{\beta f \theta} \int_0^\ell d \theta' \e^{-\beta U(\theta-\theta')- \beta f \theta'} }\hspace{10mm} \nonumber \\
&=& \int_0^\ell d \theta' \e^{-\beta V(\theta')}-(1-\e^{\beta f \ell})
\int_0^\theta d \theta' \e^{-\beta V(\theta')}.
\label{ident}
\end{eqnarray}
Using this, the expression of $\ps (\theta)$ obtained by solving the equation $\LFP \ps (\theta) = 0$ with standard techniques (see, e.g. \cite{Risken}) can be transformed into the form
\begin{equation}
\ps(\theta)=\frac{I_{-}(\theta)}{\int_0^\ell d\theta' I_-(\theta')}.
\label{psexp}
\end{equation}
The equality (\ref{ident}) also allows us to prove the relation
\begin{equation}
1 -\frac{1}{\gamma}\sum_{j=-\infty}^\infty
\frac{(\Psi_j,V')}{\lambda_j}\Phi_j'(\theta)
= \ell \frac{I_+(\theta)}{\int_0^\ell d\theta I_-(\theta)}.
\label{Qexp}
\end{equation}
(The proof of (\ref{Qexp}) is given in \ref{app:2}.)
Then, substituting (\ref{psexp}) and (\ref{Qexp}) into both (\ref{R}) and (\ref{C}) with $\omega=0$, we obtain (\ref{mud}) and (\ref{diff}). 

Before ending this section, we discuss the similarities 
and the differences between our formulation and the standard 
analysis on the basis of the Fokker-Planck equation. 
We first emphasize that calculation 
of statistical quantities by use of (\ref{norm:3}) 
is in principle equivalent 
to that using the Fokker-Planck equation that corresponds to the 
Langevin equation (\ref{eq:Langevin}).
In practice, we should use either formulation depending 
on the character of the concerned quantity. For example, 
statistics of an arbitrary function of $x(t)$ can be more easily 
calculated in the Fokker-Planck formulation. A class of quantities 
such as the multiple-time correlation functions of $x(t)-x(0)$, 
which is calculated in the Fokker-Planck formulation, 
can also be facilely calculated from (\ref{norm:3})
 by considering products of the $x$ 
variables and subsequently averaging over statistical realizations.

On the other hand, as an advantage of our formulation, 
the calculation of statistical quantities
involving $\dot x(t)$ is facilitated by use of our formulation. 
For example, as seen in the above argument,
the time correlation function of the velocity fluctuations, 
$C(t)$, can be easily calculated in our formulation, 
while such a calculation seems rather difficult, but not
impossible, in the Fokker-Planck formulation. 
Also, owing to the form of (\ref{norm:3}), it
provides a powerful tool to find a statistical quantity connected to the response function.
Recently, an illuminating example using this advantage has been demonstrated in the
reference \cite{hseq}, 
in which the steady heat flux has been expressed in terms of the violation
of fluctuation-dissipation relation.

%%%%%%%%%%%%%%%%%%%%%%%%%%%%%%%%%%
%%%%%%%%%%%%%%%%%%%%%%%%%%%%%%%%%%

\section{Concluding remarks}
\label{sec:con} 

%% generality of the method

In this paper, we presented the exact transformation of the Langevin equation (\ref{eq:Langevin}) to (\ref{norm:1}) as an example. 
Our theory may provide a novel way of understanding the ``force'' in a Langevin system as well as a new method to calculate $\tilde R(\omega)$ and $\tilde C(\omega)$ simultaneously.
Due to the general nature of the argument used in its derivation, our theory can be applied to various stochastic systems. 
For example, it is straightforward to study a B{\"u}ttiker model \cite{Butt} using our theory.
For other models, with time-dependent potentials \cite{fla, fla2, pul, pul2}, some additional techniques must be designed in order to construct a special expression that explicitly exhibits a response function.
We will report such studies in a separate paper.

The connection of the theory presented here to experimental studies is most important.
As an example, suppose that we have an experimental system consisting of a small particle exhibiting a fluctuating movement on a one-dimensional track and that the mechanics of the system is under investigation.
Then, the function $L(s)$ can be determined from the response to a time-dependent perturbation by using (\ref{LRcorrespondence}) and (\ref{resdef0}), and the statistical properties of $\zeta(t)$ can be determined from the time correlation of $\dot x(t)$ by using (\ref{zeta:def}), (\ref{M:def}) and (\ref{C}). 
Given this $L(t)$ and the statistics of $\zeta(t)$, we can write an equation of the form (\ref{norm:1}) where $\zeta(t)$ is replaced with artificial random noise satisfying the observed statistics.
Such an equation can be considered as the equation of motion for this system provided that the history dependence of $\zeta(t)$ is negligible.
Although, strictly speaking, $\zeta$ depends on the history of $x$, this description may provide a good starting point for the construction of a phenomenological theory \cite{Harada}.

The idea of re-expressing differential equations reminds us of the normal form theory, whose goal is to transform the equation in question into as ``simple'' a form as possible.
The normal form theory was extended to stochastic processes, with the main focus on bifurcation problems \cite{Edger, Coullet, Haken}.
Furthermore, mathematical techniques for random normal forms have been developed \cite{Arnold}.
However, as far as we are aware, the transformation of (\ref{eq:Langevin}) to (\ref{norm:1}) has never been presented. 

In another related work, a generalized Langevin equation 
with a memory function was derived from a nonlinear 
Langevin equation by employing a projection operator method \cite{Mori}.
In this method, functions of dynamical variables are projected 
onto a subspace of interest, by using the so-called Mori identity 
\cite{Mori65}. Then, the choice of the projection operator 
represents the essence of the problem when considering whether 
the obtained result is physically meaningful, as pointed out 
in the reference \cite{kawa}.  However, as far as we understand, 
there was no such discussion in the reference  \cite{Mori}.
Of course, whatever method we use, the important point is 
whether the obtained result is related to quantities and 
relations that can be measured experimentally.
The determination of dissipative and other forces should be done in such a way that it is clear how they can be measured experimentally.

%%%%%%%%%%%%%%%% Acknowledgment %%%%%%%%%%%%%%%%

\ack
The authors thank K. Sato and E. Knobloch for useful comments. 
This work was supported by a grant from the Ministry of Education, Science, Sports and Culture of Japan, No. 16540337.
One of the authors (TH) also appreciates the support of Research Fellowships for Young Scientists from the Japan Society for the Promotion of Science, No. 05494.

%%%%%%%%%%%%%%%% Appendix %%%%%%%%%%%%%%%%%%%%

\appendix

\section{Verifying the formula (\ref{Ldef})} 
\label{app:1}

Here, we confirm the validity of the formula (\ref{Ldef}) by directly calculating the response function from its definition (\ref{resdef0}).
To this end, we utilize a  perturbation force, $\fp(t)$, that takes the form of a step function: $\fp(t) = 0$ for $t \le 0$ and $\fp(t) = 1$ for $t > 0$.
Then, we have
\begin{equation}
\bra \dot x(t) \ket_\ve  - \vs= \ve \int_0^t ds R(s)  + O(\ve^2).
\label{r}
\end{equation}
Using (\ref{pin}), the left-hand side of this equation can be expressed as 
\begin{equation}
\bra \dot x(t)  \ket_\ve - \vs= 
\int_0^\ell d\theta p_\infty(\theta, t) 
\frac{F(\theta) + \ve \fp(t)}{\gamma} - \vs.
\label{vt}
\end{equation}
The function $p_\infty(\theta, t)$ satisfies the equation
\begin{equation}
\pder{}{t} p_\infty(\theta, t) = 
(\LFP - \frac{\ve}{\gamma} \fp(t) \pder{}{\theta}) p_\infty(\theta, t),
\label{fp}
\end{equation}
with the initial condition $p_\infty(\theta, 0) = \ps(\theta)$.

Substituting the expanded form (\ref{pinex}) into (\ref{fp}) and extracting terms proportional to $\ve$, we obtain
\begin{equation}
\pder{}{t} p^{(1)} (\theta, t) = 
\LFP p^{(1)}(\theta, t) - 
\frac{1}{\gamma} \fp(t) \ps'(\theta).
\label{fp1}
\end{equation}
Next, we write $p^{(1)}(\theta, t)$ in terms of the eigenfunctions of $\LFP$:
\begin{equation}
p^{(1)}(\theta, t) = \sum_{j=-\infty}^\infty p^{(1)}_j (t) 
\Psi_j^* (\theta).
\label{p1ex}
\end{equation}
Substituting this form into (\ref{fp1}), we  obtain
\begin{equation}
\dot p^{(1)}_j (t) 
= 
- \lambda_j p^{(1)}_j (t) 
+ \frac{1}{\gamma}\fp (t) (\Psi_0, \Phi'_j) .
\label{fp2}
\end{equation}
Integration of (\ref{fp2}) with the initial condition $p^{(1)}_j (0) =0$ yields
\begin{equation}
p^{(1)}_j (t) 
= \frac{1}{\gamma} \fp(t) (\Psi_0, \Phi'_j) \int_0^t ds \e^{-\lambda_j s}.
\label{p1j}
\end{equation}

Using the results (\ref{p1ex}) and (\ref{p1j}),
we express  (\ref{vt}) as 
\begin{eqnarray}
\bra \dot x(t)  \ket_\ve- \vs
&=& \frac{\ve}{\gamma}\left[ \fp(t) + \frac{1}{\gamma}
\sum_{j=-\infty}^\infty \fp(t) (\Psi_0, \Phi'_j) 
(\Psi_j, F) \int_0^t ds \e^{-\lambda_j s} \right] \nonumber \\
& & + O(\ve^2).
\end{eqnarray}
Comparing this with (\ref{r}), we find
\begin{equation}
\gamma R(t) = \delta(t) + \sum_{j=-\infty}^\infty \frac{1}{\gamma} 
(\Psi_j, F) \e^{-\lambda_j t} (\Psi_0, \Phi'_j).
\end{equation}
The Fourier transformation of this relation gives
\begin{equation}
\gamma \tilde R(\omega) 
= 1 + \sum_{j=-\infty}^\infty \frac{1}{\gamma} \frac{(\Psi_j, F)}{\lambda_j - i \omega} (\Psi_0, \Phi'_j),
\label{chi_final}
\end{equation}
which is found to be equivalent to (\ref{R}).
 
\section{Proof of (\ref{Qexp})}
\label{app:2}

Differentiating (\ref{Phidef}) with respect to $\theta$, we obtain
\begin{equation}
\Lambda^\ddagger \Phi_j' =-\lambda_j \Phi_j',
\end{equation}
with 
\begin{equation}
\Lambda^\ddagger \equiv -\frac{1}{\gamma}
\pder{}{\theta} V'(\theta)
+ \frac{T}{\gamma}\pdert{}{\theta}
\label{eq:tLambda}.
\end{equation}
Because $\Phi_j'$ is an eigenfunction of $\Lambda^\ddagger$, below we employ the notation
\begin{equation}
\Phi_j^\ddagger\equiv\Phi_j'.
\label{def:phidd}
\end{equation}

The operator adjoint to $\Lambda^\ddagger$, $\tLFP$, is defined through the following:
\begin{equation}
(\tLFP h_1,  h_2) \equiv (h_1, \Lambda^\ddagger h_2).
\end{equation}
The explicit form of $\tLFP$ is
\begin{equation}
\tLFP = \frac{1}{\gamma}V'(\theta)\pder{}{\theta}
+ \frac{T}{\gamma}\pdert{}{\theta}
\label{eq:tFP}.
\end{equation}
The set of eigenvalues, $-\lambda_j$, of this operator $\tLFP$ is identical to that of $\Lambda^\ddagger$.
The corresponding eigenfunctions $\Psi_j^\ddagger$ are labeled such that they satisfy the equation
\begin{equation}
\tLFP \Psi_j^\ddagger =-\lambda_j^* \Psi_j^\ddagger.
\label{defpsi}
\end{equation}
These eigenfunctions can be chosen so as to satisfy the orthogonality condition
\begin{equation}
(\Psi_i^\ddagger, \Phi_j^\ddagger)=\delta_{ij}.
\label{ons}
\end{equation}
Then, differentiating (\ref{defpsi}) with respect to $\theta$, we obtain
\begin{equation}
\LFP \Psi_j^{\ddagger'} =-\lambda_j^* \Psi_j^{\ddagger'}.
\label{defpsi2}
\end{equation}
Thus, applying (\ref{ons}),  we have 
\begin{equation}
\Psi_j^{\ddagger'}=-\Psi_j
\end{equation}
for the case $j\not =0$, and $\Psi_0^{\ddagger}=1$.

Now, we define the quantity $Q(\theta)$ as
\begin{equation}
Q(\theta)\equiv -\frac{1}{\gamma}\sum_{j=-\infty}^\infty \frac{(\Psi_j,V')}{\lambda_j}\Phi_j'(\theta).
\label{Qdef2}
\end{equation}
Using (\ref{def:phidd}), we calculate
\begin{equation}
\Lambda^\ddagger  Q= \frac{1}{\gamma}V''.
\end{equation}
Solving this equation, we obtain
\begin{equation}
Q(\theta)=-1+q_1 \e^{\beta V(\theta)}\int_0^\theta
d \theta' \e^{-\beta V(\theta')}
+q_2 \e^{\beta V(\theta)}.
\label{ares}
\end{equation}
The two constants $q_1$ and $q_2$ here are determined by the conditions
\begin{eqnarray}
Q(0) = Q(\ell), \label{PBC} \\
(\Psi_0^\ddagger, Q)=0. \label{ort2}
\end{eqnarray}

The condition (\ref{PBC}) leads to 
\begin{equation}
q_2=q_1 \frac{1}{\e^{\beta f \ell}-1}
\int_0^\ell d \theta' \e^{-\beta V(\theta')}.
\end{equation}
Substituting this into (\ref{ares}), we obtain
\begin{eqnarray}
1+ Q(\theta) &=& q_1 \frac{\e^{\beta V(\theta)}}{\e^{\beta f \ell}-1}
\left[ \int_0^\ell d \theta' \e^{-\beta V(\theta')}-(1-\e^{\beta f \ell})
\int_0^\theta d \theta' \e^{-\beta V(\theta')} \right]
\label{ares2} \nonumber \\
&=& \frac{q_1}{\e^{\beta f \ell}-1}I_{+}(\theta),
\label{ares3}
\end{eqnarray}
where we have used the identity (\ref{ident}).
Using (\ref{ares3}) in (\ref{ort2}) leads to the final form (\ref{Qexp}).

%%%%%%%%%%%%%%%%%%%%%%%%%%%%%%%%%%%%%%%%%%%%%

\end{document}